\documentclass[floatfix,aps,prl,preprintnumbers,nofootinbib,superscriptaddress,showpacs,letterpaper,twocolumn]{revtex4-1}
\usepackage{amsmath,amssymb,amsbsy,booktabs,array}
\usepackage{bm,comment}
\usepackage{rotating}
\usepackage{amsmath,amssymb,braket,slashed,mathrsfs}
\usepackage[utf8]{inputenc}

\usepackage{array}
\newcolumntype{C}{>{$}c<{$}}

\def\amulohvp{a_\mu^\text{LO-HVP}}
\def\aelllohvp{a_\ell^\text{LO-HVP}}
\def\fm{\mathrm{fm}}
\def\gev{\mathrm{Ge\kern-0.1em V}}
\def\mev{\mathrm{Me\kern-0.1em V}}
\def\reff#1{\ref{#1}}

\def\eq#1{Eq.~(\reff{#1})}

\def\fig#1{Fig.~\reff{#1}}

\def\tab#1{Table~\reff{#1}}

\newcommand{\marseille}{\affiliation{CNRS, Aix Marseille U., U. de Toulon, CPT, UMR 7332, F-13288 Marseille, France}}
\newcommand{\wuppertal}{\affiliation{Department of Physics, Bergische Universit\"at Wuppertal, Gaussstr. 20, D-42119 Wuppertal}}
\newcommand{\juelich}{\affiliation{J\"ulich Supercomputing Centre, Forschungszentrum J\"ulich, D-52425 J\"ulich, Germany}}
\newcommand{\budapest}{\affiliation{Inst. for Theor. Physics, E\"otv\"os University, P\'azm\'any P. s\'et. 1/A, H-1117 Budapest, Hungary}}

\begin{document}
\bibliographystyle{apsrev}

\title{Slope and curvature of the hadron vacuum polarization at vanishing virtuality from lattice QCD}
\author{Sz.~Borsanyi}
\wuppertal
\author{Z.~Fodor}
\wuppertal
\budapest
\juelich
\author{T.~Kawanai}
\juelich
\author{S.~Krieg}
\wuppertal
\juelich
\author{L.~Lellouch}
\marseille
\author{R.~Malak}
\marseille
\author{K.~Miura}
\marseille
\author{K.K.~Szabo}
\wuppertal
\juelich
\author{C.~Torrero}
\marseille
\author{B.~Toth}
\wuppertal
\date{\today}

\begin{abstract}

    We compute the slope and curvature, at vanishing four-momentum transfer
    squared, of the leading order hadron vacuum polarization function,
    using lattice QCD. Calculations are performed with 2+1+1 flavors
    of staggered fermions directly at the physical values of the quark
    masses and in volumes of linear extent larger than $6\,\fm$.  The
    continuum limit is carried out using six different lattice
    spacings. All connected and disconnected contributions are
    calculated, up to and including those of the charm.

\end{abstract}

\maketitle


{\em Introduction.--} The vacuum expectation value of the product of two
electromagnetic currents plays an important role in physics. It
describes how virtual particle fluctuations polarize the vacuum as it is
traversed by a propagating photon. While the contributions associated
with virtual leptons and weak bosons can be computed in perturbation theory,
those of quarks require nonperturbative methods for small photon
virtuality, because of the confinement of quarks within hadrons. Here we focus
on the latter, known as the hadron vacuum polarization or HVP.

The low energy behavior of the HVP is the limiting uncertainty in the
standard model (SM) prediction of a number of quantities. It limits
the precision with which many electroweak observables are determined
\cite{Jegerlehner:2008rs}. It also represents the leading hadronic
uncertainty in the SM prediction for the anomalous magnetic moments of
leptons, $a_\ell$ with $\ell=e,\,\mu,\,\tau$
\cite{Jegerlehner:2009ry,Miller:2012opa}. In fact, it is the limiting
factor in the SM prediction
\cite{Davier:2010nc,Hagiwara:2011af,Jegerlehner:2015stw,Jegerlehner:2009ry,Miller:2012opa}
of the much debated anomalous magnetic moment of the muon that is
currently measured to 0.54~ppm \cite{Bennett:2006fi}.

Today the HVP is best determined using dispersion relations and the
cross section of $e^+e^-$ to hadrons or the rate of hadronic $\tau$
decays
\cite{Eidelman:1995ny,Davier:2010nc,Hagiwara:2011af,Jegerlehner:2015stw}. However,
since the pioneering work of \cite{Blum:2002ii}, lattice QCD
calculations of the leading order (LO) HVP contributions, $\amulohvp$,
to $a_\mu$ have made significant progress
\cite{Aubin:2006xv,Feng:2011zk,DellaMorte:2011aa,Burger:2013jya,Chakraborty:2014mwa,Blum:2015you,Chakraborty:2015ugp,Blum:2016xpd,Blum:2013qu}. Moreover,
in the long run, this approach is likely to represent the most
cost-effective way to increase the precision of the HVP to the levels
that will soon be required by the new round of measurements of $a_\mu$
\cite{Gray:2010fp,Otani:2015jra} and, more generally, by particle physics
phenomenology.

Here we present a full lattice QCD calculation of the first two
derivatives of the HVP function at zero, euclidean virtuality. The
calculation includes all contributions from $u$, $d$, $s$ and $c$
quarks, both in their quark-connected and quark-disconnected
configurations, in the isospin limit. As shown in
\cite{Bell:1996md,deRafael:2014gxa}, the slope of the polarization
function provides an upper bound on the HVP contribution to the
anomalous magnetic moment of all three leptons. It also determines the
whole of $\aelllohvp/m_\ell^2$ in the limit that the lepton mass,
$m_\ell$, vanishes \cite{Bell:1996md}. Moreover, together with the
curvature, the slope gives $\amulohvp$ to within less than $2\%$. This
fraction, which is indicative only, derives from comparing the value
of $\amulohvp$ obtained from a simple phenomenological model to that
obtained using a $[1,1]$ Padé approximant to describe the 
virtuality dependence of the HVP function. This approximant is constructed from the slope and
curvature of the HVP function at vanishing virtuality, obtained in the same model. The
model combines the $e^+e^-$ experimental spectrum up to the $\psi'$ that is
compiled in \cite{Agashe:2014kda}, perturbative contributions above
$s=2.25\,\gev^2$ and dispersion relations. The use of Padé
approximants for determining $\amulohvp$ was first proposed in
\cite{Aubin:2012me} and was first used for a lattice QCD evaluation in
\cite{Chakraborty:2014mwa}.

\begin{figure}
    \centering
    \includegraphics{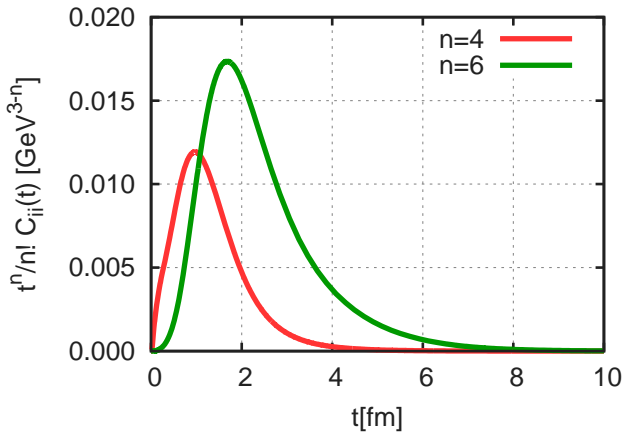}
    \caption
    {\label{fi:phenomomentkern} Kernels of two moments of the spatial
      components of the zero three-momentum, electromagnetic current
      correlator as a function of euclidean time (see \eq{eq:pi} with
      $\mu=i=1,2$ or $3$ and $\nu=0$). As described in the text, these
      moments can be used to determine the slope and curvature of the
      HVP function at vanishing virtuality. The electromagnetic current
      correlator is obtained using a phenomenological description of
      the $e^+e^-$ data compiled in \cite{Agashe:2014kda} and
      dispersion relations, as briefly described in the text.}
\end{figure}

The main challenge in calculating, on the lattice, derivatives of the
HVP function at zero virtuality is the fact that it requires
determining, with high precision, the dominant, and notoriously noisy,
$u$ and $d$ quark contributions to the electromagnetic current
correlation function at large euclidean distances. This is
particularly clear when these derivatives are obtained by computing
moments, in euclidean spacetime, of the quark, electromagnetic
two-point function, as described below, around \eq{eq:pi}. In
\fig{fi:phenomomentkern} we plot the kernels of two such
time-moments. They correspond to those needed for the first two
derivatives of the HVP function at vanishing virtuality (the higher
derivatives are obtained from higher moments). As the figure shows, the
distances that need to be reached to reliably determine the slope and
curvature are above $\sim 2$ and $\sim 4\,\fm$, respectively. Higher
derivatives require even larger distances. Here we address this
challenge by performing a high-precision calculation on lattices of
spatial extent $L\simeq 6\,\fm$ and of time extent $T$, up to
$11\,\fm$. This calculation yields a prediction of QCD for these
quantities that can be compared to present \cite{Benayoun:2016krn} and
future phenomenological determinations. We leave to a forthcoming
paper the computation of the anomalous magnetic moment of the muon.

{\em Simulations.--} We employ a tree-level improved Symanzik gauge action
\cite{Luscher:1984xn} and a fermion action for four flavors of stout-smeared
\cite{Morningstar:2003gk}, staggered quarks.  The up and down quark masses are
treated as degenerate, their ratio to the strange quark mass is tuned to the
vicinity of the physical point, which is defined from the Goldstone pion and
kaon masses.  The charm quark mass is fixed in units of the strange mass to
$m_c/m_s=11.85$ \cite{Davies:2009ih}. More information on the action together
with simulation and algorithmic details can be found in
\cite{Bellwied:2015lba}.

To set the physical mass point we use the isospin corrected pion and
kaon masses, $M_\pi= 134.8\,\mev$ and $M_K=494.2\,\mev$, from
\cite{Aoki:2016frl}. To convert the lattice results into physical
units, we use the pion decay constant $f_\pi=130.50(1)(3)(13)\,\mev$
\cite{Agashe:2014kda} which is free of electromagnetic corrections and,
to very good accuracy, equals to the decay constant in the $m_d=m_u$
limit \cite{Gasser:2010wz}. This makes our definition of the physical
point well defined in the isospin limit. In intermediate steps of the
analysis we use the Wilson-flow-based \cite{Luscher:2010iy}
$w_0$-scale \cite{Borsanyi:2012zs}.  For our finest lattice spacing
the root mean squared pion mass is about 15\% larger, than the
Goldstone pion mass.

Table \ref{ta:conf} lists the ensembles and the number of configurations used
for quark-connected and disconnected measurements. A configuration corresponds to 10
unit length Rational Hybrid Monte Carlo (RHMC)~\cite{Clark:2006fx}
trajectories. The integration over the trajectory is improved with the gradient
of the RHMC force \cite{Clark:2011ir,Yin:2011np}. The topological charge
undergoes sufficient number of tunnelings even on the finest lattices.

\begin{table}
    \begin{ruledtabular}
	\begin{tabular}{ccccc}
	    $\beta$ & $a$ $[\fm]$ & $T\times L$ & \#conf-conn & \#conf-disc\\
	    \hline
	    3.7000 & 0.134 & $ 64\times48$ & 1000 & 1000\\
	    3.7500 & 0.118 & $ 96\times56$ & 1500 & 1500\\
	    3.7753 & 0.111 & $ 84\times56$ & 1500 & 1500\\
	    3.8400 & 0.095 & $ 96\times64$ & 2500 & 1500\\
	    3.9200 & 0.078 & $128\times80$ & 3500 & 1000\\
	    4.0126 & 0.064 & $144\times96$ &  450 & -\\
	\end{tabular}
    \caption{\label{ta:conf} List of $\beta$, lattice spacings, sizes and number of configurations used
    for the connected and disconnected correlators.}
    \end{ruledtabular}
\end{table}

{\em Observables.--} The hadron vacuum polarization is derived from
the electromagnetic current $j_\mu$, which is defined as $j_\mu/e=
\frac{2}{3} \bar{u}\gamma_\mu u - \frac{1}{3} \bar{d}\gamma_\mu d -
\frac{1}{3} \bar{s}\gamma_\mu s + \frac{2}{3} \bar{c} \gamma_\mu c$,
where $e$ is the unit of electromagnetic charge. From this we build
the current-current correlator $\langle j_{\mu}(x) j_{\nu}(0)\rangle$,
in which we use the conserved lattice current at the source and
sink. No renormalization is therefore necessary.

We split up the correlator in two different ways. First $\langle j_\mu
j_\nu\rangle= \frac{e^2}{9}( 5 C_{\mu\nu}^l + C_{\mu\nu}^{s} + 4 C_{\mu\nu}^{c}
+ C^{\rm disc}_{\mu\nu})$, where the first three terms contain the connected
contractions for the light, strange and charm flavors, and the last contains
the disconnected contractions. Flavor mixing terms arise only in the latter.
We can also separate the correlator according to isospin symmetry, which is
exact in our simulations: $\langle j_\mu j_\nu\rangle= \langle j_\mu
j_\nu\rangle_{I=0} + \langle j_\mu j_\nu\rangle_{I=1}$. Here the isospin
singlet contribution is $\langle j_\mu j_\nu\rangle_{I=0}= \frac{e^2}{18}(
C_{\mu\nu}^l + 2C_{\mu\nu}^s + 8C_{\mu\nu}^c + 2C^{\rm disc}_{\mu\nu})$,
whereas the isospin triplet one is $\langle j_\mu j_\nu\rangle_{I=1}=
\frac{e^2}{2}C_{\mu\nu}^l$.  The lowest energy state contains three/two pions in
the isospin singlet/triplet channel. This fact determines the behavior of the
correlator for large separations.

We calculate the connected contributions to the correlators using
point sources.  We use the all-mode-averaging technique (AMA) of
\cite{Blum:2012uh} and 768 random source positions on each
configuration for the light quarks, 64 sources for the strange (except at $\beta=3.70$
where 128 were used) and
4 for the charm.  To compute the quark-disconnected contributions,
we apply AMA again, and exploit the approximate SU(3) flavor
symmetry on around 6000 stochastic sources \cite{Francis:2014hoa,Blum:2015you}. We use random spatial wall
sources, so only zero-momentum, time propagators are available.  For
the disconnected contribution of the charm we apply a hopping
parameter expansion. The computer time required for this entire
analysis is of the same order as the time needed for the generation of
configurations.

The $n$-th coefficient of a Taylor expansion of the vacuum polarization
scalar, $\Pi(Q^2)$, around $Q^2=0$ (i.e. $[\partial^{n}\Pi(Q^2)/(\partial
Q^2)^n]_{Q^2=0}/n!$) can be written as $\Pi_n=\frac{1}{9}( 5\Pi^l_n + \Pi^s_n
+ 4\Pi^c_n + \Pi^{\rm disc}_n)$, where each term is related to the
respective, configuration-space correlator through
moments \cite{Feng:2013xsa,Chakraborty:2014mwa}
\begin{align}
    \label{eq:pi}
    \Pi^f_{n,\mu\nu}= (-)^{n+1}\sum_x \frac{\hat{x}_\nu^{2n+2}}{(2n+2)!} C^f_{\mu\mu}(x)
\end{align}
for $f=\{l,s,c,{\rm disc}\}$, with $\nu\ne\mu$ and $\hat{x}$ defined as
$\hat{x}_\nu= \min(x_\nu,L_\nu-x_\nu)$ and where $L_\mu$ is the size of the lattice in the $\mu$-direction.  In general the result depends
on the choice of $\mu$ and $\nu$.  Three different averages, which are
invariant under spatial cubic transformations, can be constructed: one
which is an average over spatial moments, $\nu=1,2,3$, of correlators
of spatial currents, $\mu=1,2,3\ne\nu$; another, an average over
spatial moments of correlators of timelike currents,
$\mu=4$; a third, an average over time moments, $\nu=4$, of
correlators of spatial currents.  We call these averages
$\Pi_{n,ss}$, $\Pi_{n,4s}$ and $\Pi_{n,s4}$, respectively. In the
disconnected case we only have the $\Pi_{n,s4}$ average.  The averages
$\Pi_{n,ss},\Pi_{n,4s}$ and $\Pi_{n,s4}$ can be different, which is a
consequence of the finite lattice size and the asymmetry $T\neq L$. 

\begin{figure}
    \centering
    \includegraphics{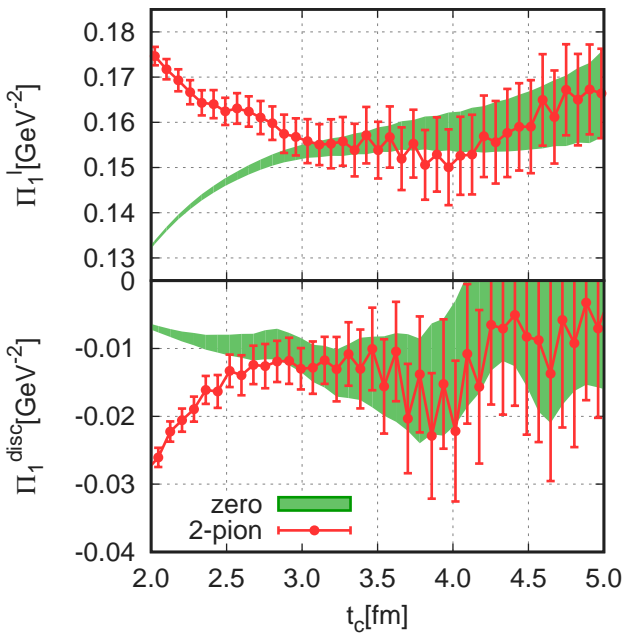}
    \caption
    {
	\label{fi:bound} Top: an upper/lower bound on $\Pi^l_1$ is obtained by
	setting the timelike correlator to a two-pion decay/zero starting at a
	separation of $t_c$. Bottom: upper/lower bound on $\Pi^{\rm disc}_1$
	is obtained by setting the correlator to zero/to the connected correlator with a two-pion decay. Results
        are for an ensemble at $\beta=3.9200$.
    }
\end{figure}
In the case of the light and disconnected correlators the signal
deteriorates quickly with increasing distance. In our analysis we
consider the spatial moments, i.e.\ $ss$ and $4s$, up to their full
extent (ca.\ $3.1\,\fm$). For the time-moment sums, $s4$, which can
extend up to ca.\ $6\,\fm$, we introduce a cut $t_c$ in time.  For
times greater than $t_c$ we replace the correlator by an upper and a
lower bound~\footnote{We are grateful to Christoph Lehner for a
  discussion of this issue.}. The connected light correlator is
proportional to the isospin triplet one, whose lowest-energy
contribution comes from a two-pion state. Therefore up to exponentially
suppressed corrections in $T$ the correlator satisfies
\begin{align}
    \label{eq:bound1}
    0 \leq C^{l}(t) \leq C^{l}(t_c)\frac{\varphi(t)}{\varphi(t_c)},
\end{align}
where $\varphi(t)= \cosh\left[E_{2\pi}(T/2-t)\right]$ and $E_{2\pi}$
is the energy of two pions, each with the smallest nonvanishing
lattice momentum, for which we use $2\pi/L$.  Typically the two bounds agree for
$t_c\gtrsim 3\,\fm$, as seen for example on the top plot of
\fig{fi:bound}. A similar conclusion is reached for the second
derivative $\Pi^{l}_{2,s4}$.  In our analyses, we take $t_c=3.1\,\fm$
on the light connected timelike correlators and average the two
bounds to get the final result.  Pion-pion interactions can change the smallest two-pion
momentum from $2\pi/L$ in that channel.  Using the model of \cite{Luscher:1991cf} and
neglecting four-pion contributions, we determine the change in the
momentum to be around 3\%. We checked, that such a reduction of the momentum
changes the result on $\Pi^{l}$ by a small fraction of the statistical
error.

The disconnected contribution alone can be constrained for large
enough time separations, where the isospin singlet channel, dominated
by three-pion states, can be neglected compared to the triplet,
dominated by two-pion ones. Here we have
\begin{align}
    \label{eq:bound2}
    0 \geq [2C^s {+} 8C^c {+} 2C^{\rm disc}](t) \geq -C^{l}(t_c)\frac{\varphi(t)}{\varphi(t_c)}
\end{align}
up to corrections exponentially suppressed in $T$.
This gives an upper and a lower bound on $\Pi^s+4\Pi^c+\Pi^{\rm
  disc}$, which can be used to determine the time $t_c$ after which the two
bounds agree within errors. At large $t$, the connected
strange and charm contributions in \eqref{eq:bound2} are exponentially suppressed, and their
presence does not make a
difference when determining $t_c$ so we neglect them.  In the bottom plot of
\fig{fi:bound} we show the upper and lower bounds on $\Pi^{\rm
  disc}$. In our analyses we average the two bounds for
$t>t_c=2.7\,\fm$.

{\em Results.--} To obtain our final results in the continuum limit and at the physical point,
we fit the lattice results to a function which depends on the pion and kaon
masses and on the lattice spacing squared $a^2$. Since the simulations were
done around the physical point, a linear pion/kaon mass dependence is always
sufficient. For the light, strange and disconnected contributions reasonable
fit qualities can be achieved with a linear $a^2$ dependence.

\begin{figure}
    \centering
    \includegraphics{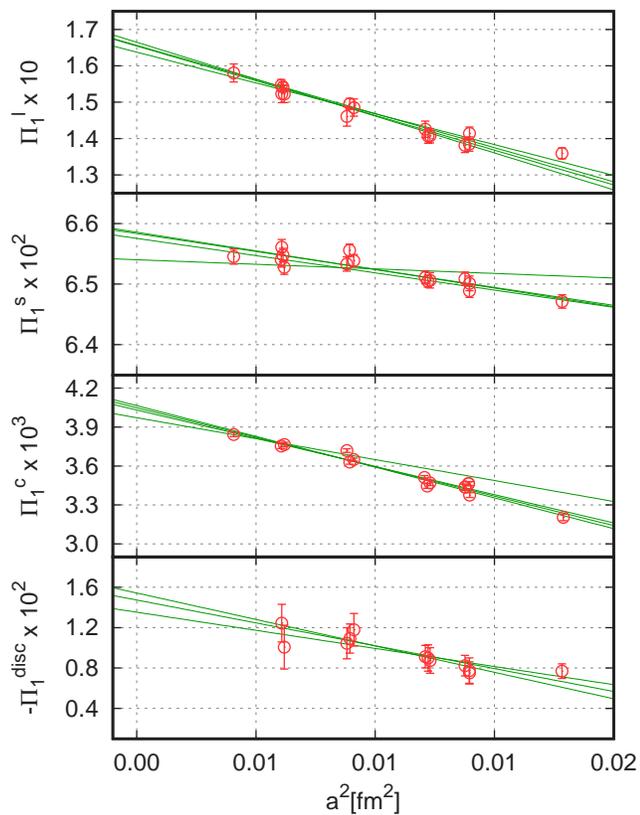}
    \caption
    {
	\label{fi:main} Continuum extrapolations of the light, strange, charm
	and disconnected contributions to $\Pi_{1}$. The lines are fits linear in $a^2$,
	the different lines correspond to different cuts in the lattice spacing.
    }
\end{figure}

\begin{table*}
    \centering
    \begin{ruledtabular}
	\begin{tabular}{lCC}
	    & \Pi_{1}[\gev^{-2}] & \Pi_{2}[\gev^{-4}]\\
	    \hline
	    light         &  0.1653(17)(16)             &  -0.295(10)(7)                \\
	    strange $\times 10^{2}$     &  6.57(1)(3)  &  -5.33(1)(4)  \\
	    charm   $\times 10^{4}$      &  40.3(2)(6)  &  -2.66(3)(11)   \\
	    disconnected  $\times 10^{2}$ &  -1.5(2)(1) &  4.4(1.0)(0.4) \\
	    \hline
	    $I=0$ & 0.0167(2)(2)   &  -0.018(1)(1)  \\
	    $I=1$ & 0.0827(8)(8)   &  -0.147(5)(4)  \\
	    total & 0.0993(10)(9)  &  -0.165(6)(4)  \\
	    \hline
	    I=1 FV corr. &0.0006(23) & -0.016(10) \\
	    \hline
	    {\bf total + FV} & \mathbf{0.0999(10)(9)(23)(13)} & \mathbf{-0.181(6)(4)(10)(2)}  \\
	\end{tabular}
    \end{ruledtabular}
    \caption
    {
	\label{ta:main} Final results on the first and second derivative of the
	hadron vacuum polarization scalar at zero squared-momentum. The first
	four lines contain the continuum extrapolated lattice results. The next
	three lines are the isospin singlet, triplet and total contributions.
	The first error is statistical and the second is systematic. In the
	following line we give an estimate of the finite-volume effects of the
	$I=1$ contribution using LO $\chi$PT. In that approach, the
	finite-volume effects on $\Pi^l/\Pi^{disc}$ are 2/-1 times those on the $I=1$
	contribution, whereas those on the $I=0$ contribution vanish. The last
	line contains the total contribution corrected for finite-volume effects,
	the third uncertainty is the one associated with this correction. The fourth
	uncertainty corresponds to isospin breaking contributions (see text).
    }
\end{table*}

We begin with the light quark contribution $\Pi^l$, paying special
attention to the difference between the averages $\Pi^l_{ss}$,
$\Pi^l_{4s}$ and $\Pi^l_{s4}$. Some ensemble show differences, but
these are not significant statistically.  If we assume no difference
between the three averages and fit all ensembles and all three
averages together in one fit, we get a reasonable fit quality
$\chi^2/{\rm dof}=45/41$. Also, if we include additional fit
parameters describing the difference between the three averages, they
come out zero within error bars. This remains true after dropping the
coarser lattices from the fit, for which the two-pion, finite-volume
effects are reduced by taste violations. Therefore we average over the
combinations $ss$, $4s$ and $s4$ in this work and leave an
investigation of differences for the future, when statistical errors
are reduced.  The upper panel of \fig{fi:main} shows the continuum
extrapolation of $\Pi^l$.  The coarsest lattice gives a value about
15\% smaller than the continuum limit. The final central value and
systematic error on the continuum limit are the mean and standard
error of the Akaike-Information-Criterion-weighted distribution
obtained by imposing four different cuts on lattice spacing (no cut,
$a\le 0.118, \; 0.111, \; 0.095\;\fm$) in the extended frequentist
approach of \cite{Durr:2008zz,Borsanyi:2014jba}. The results for the
first and the second moments are given in \tab{ta:main}.

The strange and charm quark contributions are plotted in the second and third
panels of \fig{fi:main} and the respective continuum extrapolated values
are given in \tab{ta:main}. The strange channel has much smaller lattice
artefacts than the light, since it is much less affected by taste
violations. The difference between the continuum and the coarsest lattice is
about 2\%. For the charm the fit qualities are much worse,
because the precision of the data is orders of magnitude better than for the lighter
flavors.  Here we fit only a random subset of 10 configurations, which
increases the statistical error and leads to good fit qualities. Results on the coarsest
lattice deviate by about 20\% from the continuum limit.

The lattice spacing dependence of the disconnected $\Pi_1$ is shown in the
bottom panel of \fig{fi:main}.  We calculate the charm quark contribution to
the disconnected term on the coarsest lattice: we find it to be $0.1\%$ of the
total disconnected result, i.e. much smaller than the total disconnected
statistical error. We therefore discard the charm from the disconnected term at
all lattice spacings.  $\Pi_1^{\rm disc}$ has the largest lattice artefacts
among the four contributions. Since we have one less lattice spacing available
than in the connected cases, we apply only three cuts in $\beta$. These
extrapolations are also shown in \fig{fi:main}.

The total result for $\Pi_1$ is the sum of the four contributions. It
has a combined error of about $1.4$\%. All of these results apply
to a box size of $6\,\fm$ and to the isospin symmetric case.

In the absence of a systematic study with simulations in a variety of
volumes, only model estimates of finite-volume effects can be made. As
argued in \cite{Aubin:2015rzx,Francis:2013qna}, for large volumes those effects will
be governed by pion contributions that can be computed in chiral
perturbation theory ($\chi$PT) \cite{Aubin:2015rzx}. Since the $I=0$ channel is dominated
by three-pion exchange, the finite-size effects are expected to be
smaller than those of the $I=1$ contribution, which are already
small. Thus we consider only the latter. For all three index
combinations, $ss$, $4s$ and $s4$, we calculate the difference between
the infinite and finite volume moments at one loop in $\chi$PT. We
then take the average of the maximum and the minimum differences as
our central value, with an
uncertainty given by the half distance between the maximum and
minimum. We record these corrections for $\Pi_1$ and $\Pi_2$ in 
\tab{ta:main}. For the first derivative the correction is on the
level of $2\%$, whereas on the second derivative it is of order
$10\%$. This correction increases rapidly with moment number, therefore
we have chosen not to quote moments beyond the
second one.

Concerning isospin breaking corrections, while little is known about
how they modify the slope and curvature of the hadronic vacuum
polarization function, more is known about their contribution to the
anomalous magnetic moment of the muon. Compared to phenomenological
determinations
\cite{Eidelman:1995ny,Davier:2010nc,Hagiwara:2011af,Jegerlehner:2015stw},
our $m_d=m_u$ calculation without QED is missing a number of
effects~\footnote{We are grateful to Maurice Benayoun and Fred
  Jegerlehner for very informative discussions on this
  subject.}. Noting their contributions to $\amulohvp$ in parentheses
in units of $10^{-10}$, these effects are $\rho$-$\omega$ ($2.80\pm
0.19$ from \cite{Davier:2009ag}) and $\rho$-$\gamma$ ($-2.71\pm 0.27$
from \cite{Jegerlehner:2011ti,benjeger16}) mixing, final state
radiation ($3.86\pm 0.39$ from \cite{Jegerlehner:2009ry} with a 10\%
error added), and the $\pi^0\gamma$ ($4.42\pm 0.19$ from
\cite{Davier:2010nc}) and $\eta\gamma$ ($0.64\pm 0.02$ from
\cite{Davier:2010nc}) contributions. This leads to a correction of
$(9.0\pm 0.5)\times 10^{-10}$, i.e. 1.3\% of the result for
$\amulohvp$ given in \cite{Davier:2010nc}. However, a competing effect
enters. In our calculation without electromagnetism, the charged pion
has a mass which is smaller than its physical value (see above). But a
smaller charged-pion mass has the effect of enhancing the two-pion
contribution to $\amulohvp$ and thus leads to a correction whose sign
is opposite to the correction associated with the sum of effects
discussed above. Moreover, a phenomenological description based on
$e^+e^-$ data indicates that their magnitudes are very
close \cite{benjeger16}. Thus, we assume here that the total correction which has to be
added to an isospin-limit determination of $\amulohvp$ is
$(0.0\pm 1.3)\%$, where we have taken the error to be of the typical
size of the corrections themselves. Because of the dominant role which
$\Pi_1$ plays in determining $\amulohvp$, one expects a tantamount
correction on that coefficient. Inferring the correction on $\Pi_2$ is
less direct, but we assume here that it is of the same size as for
$\Pi_1$. Thus, we add $(0.0\pm 1.3)\%$ of $\Pi_1$ and $\Pi_2$ to our
results for these quantities, after they have been corrected for
finite-volume effects. 

Putting everything together we
quote our final results for the first two moments in the last row of
\tab{ta:main}. Combining all four errors in quadrature, 
we obtain $\Pi_1$ with a total uncertainty of $2.9\%$ and $\Pi_2$ of
$7.2\%$.

{\em Discussion.--} It is interesting to compare these results with
those in the literature. The only other lattice determination of
on $\Pi_1$ and $\Pi_2$, near the physical mass point, comes from a
single, $N_f=2+1+1$ staggered simulation performed on a $(5.8^3\times
7.7)\,\fm^4$ lattice with spacing $a=\,0.12\,\fm$
\cite{Chakraborty:2014mwa,Chakraborty:2016mwy}. However, since the
action used is different from ours and a comparison with our continuum
limit results could be misleading, we choose not to exhibit such a
comparison here.

The only phenomenological determination of the slope and derivative parameters,
$\Pi_1$ and $\Pi_2$, is the recent one of \cite{Benayoun:2016krn}. Taking their
``data direct'' results, which are obtained from an interpolation of
$e^+e^-\to\mbox{hadrons}$ data, and converting them to our conventions, we get
$\Pi_1=0.0990(7)\,\gev^{-2}$ and $\Pi_2=-0.2057(16)\,\gev^{-4}$. These numbers
can be compared to our final results, i.e. those given in the last row of
\tab{ta:main}. In absolute value, their result for $\Pi_1$ is 0.3 combined
standard deviations smaller than ours and for $\Pi_2$, 1.9 $\sigma$ larger. The
latter might be due to an underestimate of FV corrections in our determination
of the second moment, or some problem with the experimental data used in the
phenomenological analysis of \cite{Benayoun:2016krn}.

{\em Acknowledgments.--} LL thanks M.~Benayoun, C.~Davies, F.~Jegerlehner,
C.~Lehner, H.~Leutwyler and E.~de Rafael for very informative
discussions. Computations were performed on JUQUEEN and JUROPA at
Forschungszentrum J\"ulich, on Turing at the Institute for Development
and Resources in Intensive Scientific Computing (IDRIS) in Orsay, on
SuperMUC at Leibniz Supercomputing Centre in M\"unchen, on Hermit at
the High Performance Computing Center in Stuttgart.  This project was
supported by in part by the OCEVU Laboratoire d’excellence
(ANR-11-LABX-0060) and the A*MIDEX Project (ANR-11-IDEX-0001-02),
which are funded by the “Investissements d’Avenir” French government
program and managed by the “Agence nationale de la recherche” (ANR),
by the Gauss Centre for Supercomputing e.V and by the GENCI-IDRIS
supercomputing grant No. 52275.

\bibliography{gmk}
\bibliographystyle{apsrev4-1}

\end{document}